\documentclass{article}

\usepackage{amsmath,amstext,amsthm,enumerate,enumerate}
\usepackage[dvips]{graphicx}

\def\a{{\alpha}}

\def\b{{\beta}}
\def\g{{\gamma}}
\def\lam{{\lambda}}

\def\w{{\omega}}
\def\h{\hbar}

\theoremstyle{plain}

\theoremstyle{definition}

\begin{document}

\title{Hamiltonians separable in cartesian coordinates and third-order integrals of motion}

 \author{Simon Gravel\\\textit{\small D\'epartement de physique et Centre de recherche math\'ematiques}\\
 \textit{\small Universit\'e de Montr\'eal, C.P.\-6128, Succursale Centre-Ville} \\
 \textit{\small Montr\'eal, Qu\'ebec} \\ \textit{\small H3C 3J7} \\ \textit{\small Canada}\\\small
 Simon.Gravel@UMontreal.CA}

\maketitle
\begin{abstract}
We present in this article all Hamiltonian systems in $E(2)$ that
are separable in cartesian coordinates and that admit a
third-order integral, both in quantum and in classical mechanics.
Many of these superintegrable systems are new, and it is seen that
there exists a relation between quantum superintegrable
potentials, invariant solutions of the Korteweg-De Vries equation
and the Painlev\'e transcendents.
\end{abstract}

\pagebreak

\section{Introduction}
In classical mechanics, an n-dimensional Hamiltonian system is
called Liouville integrable if it allows $n$ functionally
independent integrals of motion in involution (including the
Hamiltonian), that is

\begin{equation*}
\begin{split}
\{H,X_i\}&=0,\\
\{X_i,X_j\}&=0,  \forall i,j.
\end{split}
\end{equation*}

The Hamiltonian $H=H(x_1,...x_n,p_1,...,p_n)$ and the integrals of
motion $X_i=X_i(x_1,...x_n,p_1,...,p_n)$ must be well defined
functions on phase space~(\cite{Ar,Go}). The system is
superintegrable if it allows more than $n$ functionally
independent integrals, $n$ of them in involution. It is called
maximally superintegrable if it allows $2n-1$ integrals of motion.
The best known superintegrable systems in $n$ dimensions are the
harmonic oscillator $V=\omega r^2$ and the Coulomb potential
$V=\frac{\a}{r}$, and they are indeed maximally superintegrable.
This may be closely related to Bertrand's theorem (\cite{Ar, Be})
which states that these are the only rotationally invariant
systems for which all finite trajectories are closed.

In quantum mechanics, a Hamiltonian system is said to be
integrable if there exists a set $\{X_i\}$ of $n$ well defined,
algebraically independent operators (including the Hamiltonian)
that commute pairwise. It is superintegrable if it possesses
further independent operators, $\{Y_j\}$ that commute with the
Hamiltonian. The $Y_j$ do not  necessarily commute with each
other, nor with the $X_i$.

The independence of operators in quantum mechanics remains to be
defined rigorously \cite{GW,Ht1, Ht3,We}. Since we are dealing
here only with polynomial differential operators, we can proceed
by analogy with the classical case, keeping in mind that a
rigorous definition will be needed as soon as we will want to make
some more general statements. This choice of a definition will be
used only for discussion purposes, since we will find all
potentials that admit third-order integrals and all their
integrals. The results obtained will therefore hold for any
definition of the independence of operators.

Integrable and superintegrable systems, both in quantum and in
classical mechanics, attracted considerable interest in the last
years. Extensive literature exists about systems with second-order
integrals of motion, either in euclidian space
\cite{Ev1,Ev2,Ev3,FMSUW,MSVW,WSF}, or in spaces with nonzero
constant \cite{KMP, RS2} or nonconstant curvature \cite{KKW}. As
long as there was no magnetic field in the Hamiltonian, the
quantum and classical integrals of motion obeyed the same
determining equations, and therefore quantum and classical
integrability were very similar. Both properties were related to
separation of the Hamilton-Jacobi or Schr\"odinger equations, and
also to exact solvability \cite{TTW} and generalized symmetries
\cite{STW}.

Systems with higher-order integrals have been studied and
classified as early as 1935 in a well-known paper by Drach. This
paper considered classical Hamiltonians in complex Euclidian
space. Efforts were made recently to understand and classify more
completely systems with higher-order integrals in classical
\cite{GW,Ts} and quantum mechanics \cite{GW}. In spite of these
efforts still relatively few such systems are known.
 This paper is the logical sequel of a systematic search
for superintegrable systems with higher order integrals started in
\cite{GW}. Here we consider two-dimensional real Euclidian space
with a one-particle Hamiltonian;

\begin{equation*}
H=\frac{1}{2}\left(p_x^2+p_y^2\right)+V(x,y).
\end{equation*}

We request the existence of two additional integrals of motion,
one of second order in the momenta and the other of third order.

The condition of existence for second-order integral implies, both
in classical and quantum mechanics, that the Hamiltonian be
separable in cartesian, polar, parabolic or elliptic coordinates.
In this paper we consider potentials that are separable in
cartesian coordinates;

\begin{equation*}
H=\frac{1}{2}\left(p_x^2+p_y^2\right)+V_1(x)+V_2(y).
\end{equation*}

We found all such systems which admit third-order integrals.
Quantum and classical mechanics will be treated simultaneously,
for the conditions of existence of integrals of motion, even
though not equivalent, are quite similar in both cases.

\section{Existence of a third-order integral}

In quantum and classical mechanics, the general third-order
commuting operator

\begin{equation*}
X = \sum_{i+j=0}^{3} P_{ij}(x,y)p_x^ip_y^j
\end{equation*}

\noindent can be reduced to a much simpler form,

\begin{equation}
\begin{split}
X=&\sum_{\substack{i,j,k\\i+j+k=3}}A_{ijk}
\{L_3^i,p_x^jp_y^k\}+\{g_1(x,y), p_x\}+\{g_2(x,y), p_y\}\\
L_3=&xp_y-yp_x.
\end{split}
\label{opnorm}
\end{equation}

\noindent where the $A_{ijk}$ are real constants, and the $g_i$
real functions. The bracket is the anticommutator. It is not
needed in classical mechanics, but in quantum mechanics it allows
us to get rid of terms with even powers of the $p_i$ and to make
sure the operator is self-adjoint. Furthermore, its use allows us
to see clearly the relations between the quantum and the classical
case. Indeed, it was found in \cite{GW} that the requirement that
the operator commutes (or Poisson-commutes) with the Hamiltonian

\begin{equation*}
H=\frac{1}{2}\left(p_x^2+p_y^2\right)+V(x,y).
\end{equation*}
\noindent implies equations that behave well in the classical
limit. Namely, commutativity implies

\begin{align}
\begin{split}
0 &= g_1V_x +g_2V_y-\frac{\h^2}{4}\Big(f_1 V_{xxx}
+f_2V_{xxy}+f_3V_{xyy}+f_4V_{yyy}\\&~~~~~~+8A_{300}(xV_y-yV_x)
+2\left(A_{210}V_x +A_{201}V_y\right)\Big),
\end{split}\label{eqquant1}\\
(g_1)_x &= 3 f_1(y) V_x +f_2(x,y) V_y,\label{eqquant2}\\
(g_2)_y &= f_3(x,y) V_x +3 f_4(x) V_y,\label{eqquant3}\\
(g_1)_y +(g_2)_x &= 2 \left(f_2(x,y) V_x+f_3(x,y) V_y \right),
\label{eqquant4}
\end{align}

\noindent in quantum mechanics, where

\begin{align*}
f_1(y) &=- A_{300} y^3 +A_{210}y^2-A_{120}y +A_{030},\\
f_2(x,y) &=3A_{300}xy^2- 2A_{210}xy +A_{201}y^2 +A_{120}x-A_{111}y+A_{021},\\
f_3(x,y) &=-3A_{300}x^2y+A_{210}x^2 -2 A_{201}xy +A_{111} x-A_{102}y+A_{012},\\
f_4(x) &=A_{300} x^3+A_{201} x^2+A_{102} x +A_{003}.
\end{align*}

The equations in classical mechanics are obtained by setting $\h
=0$. We may notice from the quantum equations, or directly from
the condition $[H,X]=0$, that we can express all quantum
integrable potentials as $\h^2 \tilde V(x,y)$ where $\tilde V$
does not depend on $\h$. It is often more natural and interesting
though to choose arbitrary parameters contained in $\tilde V$ to
be depending on $\h$, so that the potential $V$ does not vanish in
the classical limit. One may always verify though that through an
appropriate transformation of the arbitrary parameters one can
write the potential as being proportional to $\h^2$

 The three last equations, identical in the quantum and
classical cases, yield a linear compatibility condition for $V$,
which reads
\begin{equation}
\label{compatlin}
\begin{split}
0=&-f_3 V_{xxx}+\left(2 f_2-3 f_4\right) V_{xxy}+\left(-3 f_1+2 f_3\right) V_{xyy}-f_2 V_{yyy}\\
&+2\left( f_{2y}- f_{3x}\right) V_{xx}+2\left(-3f_{1y}+f_{2x}+f_{3y}-3f_{4x}\right)V_{xy}+2\left(-f_{2y}+f_{3x}\right)V_{yy}\\
&+\left(-3f_{1yy}+2f_{2xy}-f_{3xx}\right)
V_x+\left(-f_{2yy}+2f_{3xy}-3f_{4xx}\right) V_y.
\end{split}
\end{equation}

Further nonlinear compatibility conditions can be obtained from
\eqref{eqquant1} to \eqref{eqquant4} for the potential, and these
are listed in \cite{GW}. They are quite complicated though and
were not used for the results stated in  this paper.

\section{Potentials separable in cartesian coordinates}

If we set $V=V_1(x)+V_2(y)$ in equations \eqref{eqquant1} to
\eqref{eqquant4}, we find
\begin{align}
\begin{split}
0 &= g_1V_{1x} +g_2V_{2y}-\frac{\h^2}{4}\Big(f_1 V_{1xxx}
+f_4V_{2yyy}\\&~~~~~~+8A_{300}(xV_{2y}-yV_{1x})
+2\left(A_{210}V_{1x} +A_{201}V_{2y}\right)\Big),
\end{split}\label{eqcart1}\\
(g_1)_x &= 3 f_1(y) V_{1x} +f_2(x,y) V_{2y},\label{eqcart2}\\
(g_2)_y &= f_3(x,y) V_{1x} +3 f_4(x) V_{2y},\label{eqcart3}\\
(g_1)_y +(g_2)_x &= 2 \left(f_2(x,y) V_{1x}+f_3(x,y) V_{2y}
\right), \label{eqcart4}
\end{align}

\noindent with $\h=0$ in the classical case. Equations
\eqref{eqcart2} and \eqref{eqcart3} are readily integrated, so in
the cartesian case two equations remain to be solved.

The compatibility condition \eqref{compatlin} allows us to find
ODEs for $V_1$ and $V_2$. If we set alternatively $y=0$ and $x=0$,
we find
\begin{gather}
(A_{210}x^2+A_{111}x+A_{012})V_1^{(3)}(x)+4(2 A_{210} x+A_{111})V_1''(x)+12 A_{210}V_1'(x)=a x+b\label{compxy1}\\
(A_{201}y^2-A_{111}y+A_{021})V_2^{(3)}(y)+4(2 A_{201}
y-A_{111})V_2''(y)+12 A_{201}V_2'(y)=cy+d\label{compxy2}
\end{gather}

The solutions to the homogeneous part of these equations are
easily found and brought to a simple form by translations in $x$
and $y$. If we take the first one for definiteness, we have four
different types of solution. When $A_{210} \neq 0$, we have two
possible solutions,
\begin{equation*}
\begin{split}
V_{1hom} &=\frac{c_1}{(x+\a)^2}+\frac{c_2}{(x-\a)^2}\\
V_{1hom} &= \frac{c_1}{x^2}+\frac{c_2}{x^3}.
\end{split}
\end{equation*}

If $A_{210} = 0$ and $A_{111} \neq 0$ we get
\begin{equation*}
V_{1hom} = \frac{c_1}{x^2}+c_2 x.
\end{equation*}

Finally if only $A_{012} \neq 0$, the solution may be brought to
the form

\begin{equation*}
V_{1hom}=c_2 x^2+ c_1 x.
\end{equation*}

Special solutions are also simple. If $A_{210} \neq 0$, we have
$V_{1part}=\a x^2+\b x$. Otherwise, when $A_{111} \neq 0$,
$V_{1part}=\a x^3+\b x^2$, finally, if only $A_{012} \neq 0$,
$V_{1part}=\a x^4+\b x^3$. Provided that \eqref{compxy1} or
\eqref{compxy2} do not vanish trivially, we can choose $V_1$ or
$V_2$, respectively, amongst the following functions:
\begin{enumerate}[(\text{A.}1)]
\item $f_{1} =\frac{c_1}{(x+\a)^2}+\frac{c_2}{(x-\a)^2}+c_3x^2+c_4
x$\label{f1}
\item $f_{2} = \frac{c_1}{x^2}+\frac{c_2}{x^3}+c_3x^2+c_4 x$
\item $f_3 = \frac{c_1}{x^2}+c_2 x^3+c_3x^2+c_4 x$
\item $f_4 =c_1 x^4+c_2 x^2+c_3 x$
\item $f_5 =c_1 x^3 +c_2 x$
\item $f_6 = c_1 x^2$
\item $f_7 = c_1 x$\label{f7},
\end{enumerate}

\noindent and then solve \eqref{eqcart1} to \eqref{eqcart4}. These
long but rather straightforward calculations yield the $15$
superintegrable potentials included in Table 1. Some of them are
obviously particular cases of others, but we listed them
separately to account for their additional integrals. Only the
third-order integrals are listed, some of them being trivial
consequences of lower-order ones. With the exception of the
harmonic oscillator, potentials that have first-order integrals
are not listed here for they were already presented in \cite{GW}
with all their third-order integrals. The complete integrals of
motion can be found in Appendix I.

%
%
%
%
%
%
%
%
%
%
%
%
%
%
%
%
%
%
%
%
%
%
%
%
%
%

\begin{table}[t]
\begin{tabular}{|r@{}l|c|}
\hline
 &Superintegrable potentials& Leading-order terms\\
 &&of the integrals\\
 \hline
\hline
 $\star V_a=$&$a (x^2 +y^2)$ & $L^3; \{L,p_x
 p_y\};\;\{L,p_y^2\};\;\{L,p_x^2\}$\\
\hline
 $\star V_b=$&$a (x^2 +y^2)+\frac{b}{x^2}+\frac{c}{y^2}$&$
\{L,p_xp_y\}$\label{class1}\\
\hline
 $V_c=$&$a (x^2
+y^2)+\frac{\h^2}{x^2}+\frac{\h^2}{y^2}$&$ L^3;\;
\{L,p_xp_y\}$\label{harmq1}\\

\hline $V_d=$&$a (x^2 +y^2)+\frac{\h^2}{y^2}$&$L^3;\;\{L,p_x
p_y\};\;\{L,p_y^2\}$\label{harmq2}\\

\hline
 $V_{e}=$&$\frac{\h^2}{8\a^4}(x^2
+y^2)+\frac{\h^2}{(x-\a)^2}+\frac{\h^2}{(x+\a)^2}$&$2L^3-3\a^2
\{L,p_y^2\};\; \{L,p_x^2\}$\label{harmq3}\\

\hline
 $V_{f}=$&$\frac{\h^2}{8 \a^4}(x^2
+y^2)+\frac{\h^2}{y^2}+\frac{\h^2}{(x+\a)^2}+\frac{\h^2}{(x-\a)^2}$&$
2L^3-3\a^2\{L,p_y^2\}$\label{harmq4}\\

\hline
  $V_{g}=$&$\frac{\h^2}{8 \a^4}(x^2
+y^2)$&$
2 L^3-3\a^2(\{L,p_x^2\}+\{L,p_y^2\})$\\&$+\frac{\h^2}{(y-\a)^2}+\frac{\h^2}{(x-\a)^2}+\frac{\h^2}{(y+\a)^2}+\frac{\h^2}{(x+\a)^2}$&\label{harmq5}\\

\hline
 $\star V_h=$&$a (4 x^2 +y^2)+\frac{b}{y^2}+c x$&$p_x p_y^2$\label{class2}\\

\hline
 $\star\; V_i=$&$a (9 x^2 +y^2)$&$\{L,p_y^2\}$\label{class3}\\

\hline
 $V_j=$&$a (9 x^2 +y^2)+\frac{\h^2}{y^2}$&$\{L,p_y^2\}$\\

\hline $V_{k}=$&$ \frac{\h^2}{8\a^4}(9 x^2
+y^2)+\frac{\h^2}{(y+\a)^2}+\frac{\h^2}{(y-\a)^2}$&$\{L,p_y^2\}$\\

\hline
$V_l=$&$\frac{\h^2}{x^2}+\frac{a}{y^2}$&$\{L^2,p_x\};\;\{L,p_xp_y\};\;
p_x^3$\label{potq}\\

\hline $V_m=$&$\frac{\h^2}{x^2}+\frac{\h^2}{y^2}$&$
 L^3;\;\{L^2,p_x\};\;\{L^2,p_y\}$\\&&$ \{L,p_xp_y\};\;p_x^3;\;p_y^3$\label{potr}\\

\hline
 $V_n=$&$ a x +\frac{\h^2}{y^2}$&$\{L,p_y^2\};\;
p_y^3;\;p_xp_y^2$\\

\hline
 $V_o=$&$\frac{\h^2}{y^2}+ V(x)$&$p_y^3$\\
 \hline

\end{tabular}
\caption{Superintegrable potentials that satisfy linear
compatibility conditions for nonzero parameters.} \label{table1}
\end{table}

Many of these potentials were not known. The only classical
potentials among these are indicated with a $\star$. These are
well-known superintegrable potentials (see e.g. \cite{FMSUW}), and
all of them, except $V_i$, are in fact quadratically
superintegrable.

All the potentials are superintegrable in the quantum case. We
therefore notice that classical nontrivial potentials can have
many different quantum equivalents. The classical harmonic
oscillator $V_a$ can be seen as a limiting case of the quantum
potentials $V_a$, $V_c$ and $V_d$, and also, if we set
$\a=\sqrt{\h}/\omega$, of $V_e$, $V_f$ and $V_g$, not to mention
the similar potentials that can be obtained by permutations of $x$
and $y$. The anisotropic harmonic oscillator with ratio $1:3$ also
admits many quantum deformations but, interestingly, the
anisotropic oscillator with ratio $1:2$ does not admit such
deformations. Notice also that if we want to deal with real
potentials only, $\a$ must be either real or purely imaginary in
potentials $V_e$, $V_f$, $V_g$ and $V_k$. Therefore these have as
a classical limit harmonic oscillators with $a>0$.

 All quantum superintegrable potentials reduce to classical ones when the
classical limit is considered, sometimes in more than one way. For
example, potentials $V_e$, $V_f$ and $V_g$ give the free motion
potential instead of the harmonic oscillator if $\a$ remains
constant as $\h \rightarrow 0$.

These potentials all satisfy the linear equations \eqref{compxy1}
and \eqref{compxy2}, and can be expressed as sums of simple
superintegrable potentials.

 Let us now set $A_{210}=A_{111}=A_{012}=0$ so that \eqref{compxy1} vanishes
trivially. We may also assume that $V_1$ does not take one of the
forms (A.\ref{f1}) to (A.\ref{f7}), for we have already worked
these cases out. This is quite useful, for if we set $y=1$ in
\eqref{compatlin}, we obtain for $V_1$ an equation of the same
form as \eqref{compxy1} with different coefficients. These
coefficients must therefore vanish, so
$A_{300}=A_{201}=A_{102}=0$. This is a significant simplification
that allows us to restrict our attention, when considering
equation \eqref{compxy2}, to the following three cases:
\begin{enumerate}[i)]
\item $V_2=a y^2;\label{ay2}$
\item $V_2=ay;\label{ay1}$
\item $A_{120}=A_{021}=0\label{qcque}$
\end{enumerate}

Before we consider each case separately, it is worth noticing that
potentials of the form $V=V_1(x)$ that admit third-order integrals
independent of $y$ and $p_y$ should appear as solutions here, for
the integral remains if we add a function $V_2(y)$ to these
potentials. These potentials were found in \cite{GW} and
\cite{Ht3} to satisfy equation
\begin{equation}
\label{Weierstrass} \h^2 V_{1}'^2=4V_1^3-g_2V_1-g_3,
\end{equation}

\noindent and can therefore be written as

\begin{equation}\label{Weiersol}
V_1=\h^2 \mathcal{P} (x),
\end{equation}

\noindent where $\mathcal{P} (x)$ is the Weierstrass elliptic
function. Since the $y$ variable plays no role here and these
potentials admit integrals with leading-order terms proportional
to $p_x^3$, these solutions will appear in cases \eqref{ay2} to
\eqref{qcque}.

\subsection{Case \ref{ay2}: $V=V_1(x)+ay^2$}

When $V=V_1(x)+ay^2$ and $a \neq 0$, we find that $A_{300}=0$, and the following two equations must be satisfied:
\begin{align}
0&=A_{030} \left(\h^2V_1^{(3)}-6 (V_1^2)'\right)+\gamma_1 V_1'\label{vay21}\\
\begin{split}
0&=A_{120}\left(-\h^2V_1^{(4)} -24 a(xV_1)'+6(V_1^2)''-4 a x^2
V_1'' +8a^2 x^2\right) \\&~~+8a
A_{021}\left(2ax-(xV_1')'-2V_1'\right)+4 \eta(2 a-V_1'').\label{vay22}
\end{split}
\end{align}

\noindent where $\gamma_1$ and $\eta$ are arbitrary constants.
When $A_{030} \neq 0$, equation \eqref{vay21} is equivalent to
\eqref{Weierstrass} (up to a translation of $V_1$ to get rid of
$\gamma_1$), hence its solutions are of the form \eqref{Weiersol}.
These potentials cannot satisfy simultaneously equation
\eqref{vay22} for nontrivial parameters. This can be observed by
expanding \eqref{Weierstrass} in series around $x=0$ and
substituting the result in \eqref{vay22}. Therefore solutions
given by $A_{030} \neq 0$ are of no special interest here.

Let us now set $A_{030}=\gamma_1=0$. Equation \eqref{vay22} can be
greatly simplified. We assume that $A_{120} \neq 0$, for otherwise
equation \eqref{vay22} can be solved to give potential
\eqref{class2}. Then by an appropriate translation of $x$ and $V$,
we can get rid of the terms involving $A_{021}$ and $\eta$ and
finally divide by $A_{120}$;

\begin{equation}\label{ord4}
0=-\h^2V_1^{(4)} -24 a(xV_1)'+6(V_1^2)''-4 a x^2 V_1'' +8a^2 x^2.
\end{equation}

This equation admits a first integral, namely

\begin{equation}\label{int3}
k=\h^2\left(x V_1'''-V_1''\right)+4x\left(a
x^2-3V_1\right)V_1'+6V_1^2+12 a x^2 V_1-2 a^2 x^4.
\end{equation}

Both \eqref{ord4} and \eqref{int3} can be simplified by setting
$V_1=W(x)+a x^2/3$. Then

\begin{equation}\label{F-1}
\h^2 W^{(4)}=12WW''+12 (W')^2+b x W'+ 2b W-\frac{1}{6}b^2 x^2
\end{equation}
\noindent with $b=-8a\neq0$ for \eqref{ord4}, and
\begin{equation}\label{quanty2}
k_2 =3 \h^2\left(x W'''-W''\right)-18 x (W^2)'+2 (2 a x^2+3 W)^2.
\end{equation}
\noindent for the first integral. Equation \eqref{F-1} is well
known. It is equivalent to equations (3.16) in \cite{CK} and
(2.17) in \cite{LW}, which were obtained by nonclassical reduction
of the Boussinesq Equation. It was also shown in \cite{CW} to be a
nonclassical reduction of the Kadomtsev-Petviashvili equation. It
has the Painlev\'e property, and, when $b\neq 0$, its solution,
given in \cite{Co} (equation 2.88) may be written in terms of the
fourth transcendant function of Painlev\'e, namely

\begin{equation}
W=\frac{\h}{2}b_1 P_4'(x,\frac{b}{\h^2})-\frac{1}{2}b P_4^2(x,\frac{b}{\h^2})-\frac{1}{2}b x
P_4(x,\frac{b}{\h^2})-\frac{1}{6}(\frac{b}{2} x^2+\h^2 K_1-\h b_1)
\end{equation}

\noindent where $b_1 \equiv \pm \sqrt{-b}=\pm \sqrt{8a}$ and
$P_4(x,\frac{b}{\h^2})=P_4(x,\frac{b}{\h^2},K_1,K_2)$ is the fourth transcendant function
of Painlev\'e, and therefore satisfies equation

\begin{equation}
\begin{split}
P_4''(x,\a)=&\frac{(P_4'(x,\a))^2}{2 P_4(x,\a)}-\frac{3 \a}{2}
P_4(x,\a)^3-2\a x P_4(x,\a)^2\\&-(\frac{\a}{2} x^2+K_1)
P_4(x,\a)+\frac{K_2}{P_4(x,\a)}.
\end{split}
\end{equation}

$K_1$ and $K_2$ are integration constants. The potential therefore
reads
\begin{equation}
\begin{split}
V(x,y)=&a (x^2+y^2)+\frac{\h}{2}b_1 P_4'(x,\frac{-8a}{\h^2})+4a
P_4^2(x,\frac{-8a}{\h^2})\\&+4 a x P_4(x,\frac{-8
a}{\h^2})+\frac{1}{6}(-\h^2 K_1+\h b_1).
\end{split}
\end{equation}

This potential admits as special cases two anisotropic harmonic
oscillators, $V=a (x^2+y^2)$ when $K_2=0$ (and $P_4=0$), and
$V=a(x^2/9+y^2)$ when $K_1=0$ and $K_2=-1/18$ (and $P_4=-x/3$),
as well as all their quantum deformations that have the form $V=a(p^2 x^2+y^2)+f(x)$, that is potentials $V_d$, $V_e$, $V_j$ and $V_k$ (up to a permutation of $x$ and $y$).

The constant term in the potential, $(-\h^2 K_1+\h b_1)$, can be
set to zero, but we will keep it in order to be able to write the
quantum and classical integrals in a unified way.

In classical mechanics, the equation \eqref{int3} with $\h=0$
admits a first integral, which reads

\begin{equation*}
c =\frac{(9V_1-ax^2)(V_1-ax^2)^3+ \frac{k^2}{4}-k(V_1-a
x^2)(3V_1+ax^2)}{x^2}.
\end{equation*}

 We may therefore write the solution for $V_1$ implicitly as

\begin{equation}\label{implic1}
c x^2- d^2+2 d(V_1-a x^2)(3V_1+ax^2)=(9V_1-ax^2)(V_1-ax^2)^3,
\end{equation}
\noindent where $c$ and $d$ are arbitrary constants. If $c=d=0$,
we find either the familiar anisotropic harmonic oscillators, or a
potential obtained by joining at $x=0$ two halves of anisotropic
harmonic oscillators with different ratios. Even though this
potential does not have a continuous second derivative, it can be
obtained as a limiting case of the family of smooth
superintegrable potentials \eqref{classnouv2}.

In the general case the potential may be expressed as the root of
a fourth-order polynomial with three arbitrary parameters ($a$,
$c$ and $d$), although we can set $a=1$ and one of the other two
coefficients to $\pm 1$ by scaling $x$ and $V_1$.

Since equation \eqref{implic1} may describe new two-dimensional
potentials with bounded motion, it is worth studying for
interesting special cases. Indeed, if we assume $a>0$ and
$d\geq0$, we can consider the case $c=128 a^4 \tilde{d}^3/27^2$
and $d=4 a^2 \tilde d^2/27$. Four solutions exist in that case,
two of which are translated harmonic oscillators. The remaining
two potentials are

\begin{equation}\label{classnouv}
V=ay^2 +\frac{a}{9}\left(2 \tilde d+5 x^2\pm 4x  \sqrt{ (\tilde d
+x^2)}\right).
\end{equation}

Up to an additive constant, potentials \eqref{classnouv} are equal
to

\begin{equation}\label{classnouv2}
V=ay^2+\frac{a}{9}\left(x \pm 2 \sqrt{\tilde d+x^2}\right)^2.
\end{equation}

Those potentials are smooth interpolations between anisotropic
harmonic oscillators with ratio $1:1$ and $1:3$. When $\tilde d
=0$ (and therefore $c=d=0$) they are the junctions of the two
halves of harmonic oscillators mentioned above.

The integral of motion is similar in quantum and classical
mechanics, and reads

\begin{equation}
X=\{L,p_x^2\}+\{ax^2y-3y
V_1(x),p_x\}-\frac{1}{2a}\{\frac{\h^2}{4}V_{1xxx}+(ax^2-3V_{1})
V_{1x} ,p_y\},
\end{equation}

\noindent where $V_1$ is a solution to equation \eqref{implic1}.
Therefore the integral for \eqref{classnouv2} must be slightly
modified to take into account the constant term we removed.

\subsection{Case \ref{ay1}: $V=a y+V_1(x)$} Here we find again two
equations:
\begin{equation}
\begin{aligned}
0&=\h^2 A_{120} V_1^{(3)}-6A_{120}(V_1^2)'+12a^2 A_{003}+\gamma_1 V_1' ,\\
0&=\frac{A_{030}}{4}\left(-6(V_1^2)'+\h^2 V_1^{(3)}\right)'-\frac{a}{2}A_{120}\left(6 (xV_1)'+x^2 V_1''\right)\\
&~~-a A_{021}\left((xV_1')'+2V_1'\right)-\gamma_2 V_1''+a
\gamma_1.
\end{aligned}
\end{equation}

This time we cannot treat the equations separately, but we can,
when $A_{120} \neq 0$, translate $V$ to annihilate $\g_1$. Then we
can substitute the first equation in the second to get rid of
$A_{030}$, and finally translate $x$ to get rid of $A_{021}$. We
can then solve the remaining system, first by solving the second,
linear equation, and then by substituting the result in the first
one. The only solution remaining is then potential $V_n=a
y+\frac{\h^2}{x^2}$. Hence we can set $A_{120}=0$, and therefore
also $A_{003}=\g_1=0$ (as $V_1\neq b x$ and $a\neq 0$). In the
second equation we can set $\gamma_2=0$. If then $A_{030}=0$, we
find $V_1=a/x^2$ which was already classified. If $A_{030}\neq 0$
and  $A_{021}=0$, though, the potential $V_1(x)$ is solution to
\begin{equation}\label{painleve1}
\h^2 V_1''=6V_1^2+\lambda x+k
\end{equation}

If $\lambda \neq 0$ we can set $k=0$, and we find that $V_1$ can
be expressed in terms of the first Painlev\'e transcendent:

Comme $\lam$ est une constante arbitraire, il est plus pratique
d'écrire $\w^5= \frac{\lam}{\h^4},$

\begin{equation}
V=a y+\h^2 \w^2P_1\left(\w x\right),
\end{equation}

\noindent with $\w^5= \frac{\lam}{\h^4}.$

The integral of motion is

$$X=2p_x^3+3\{V_2(x),p_x\}+\{\frac{\h^4 \w^5}{4
a}, p_y\}.$$

Notice that in order to consider the limiting case $a=0$, we can
multiply the integral by $a$, and thus we find a potential that
depends only on $x$ with a trivial $p_y$ integral. We can also set
$\w^5=a {\w^5}'$ first, in which case we find the potentials
\eqref{Weiersol}. If we look for the classical limit, we find

\begin{equation}
V=ay+b\sqrt{x}.
\end{equation}

with the integral
$$X=2 p_x^3+3 b \{\sqrt{x},p_x\}-\{\frac{3 b^2}{2 a}, p_y\}.$$

Returning to equation \eqref{painleve1} and assuming $\lam=0$, we
find again that $V_1$ has the form \eqref{Weiersol}, and the
integral depends only on $x$ and $p_x$.

Finally, if we set $A_{030}A_{021}\neq 0$, we have to solve the
equation
\begin{equation}
0=\left(-\frac{3}{2}(V_1^2)'+\frac{\h^2}{4} V_1^{(3)}\right)'+b
\left((xV_1')'+2V_1'\right),\;\:b=\frac{-a A_{021}}{A_{030}}\neq
0.
\end{equation}
It can be integrated once to give
\begin{equation*}\label{eqc1}
C_1=-\frac{3}{2} (V_1^2)'+\frac{\h^2}{4} V_1^{(3)}+b
\left((xV_1')+2V_1\right).
\end{equation*}

We can set $C_1=0$ by translations in $x$ and $V$. The equation
admits a first integral which reads
\begin{equation}\label{intprem2}
2 b \hbar^2 \left(V_1(x)-b x\right)V_1''(x)+b \hbar^2\left(2 b
-V_1'(x)\right)V_1'(x)-8 b V_1(x) \left(V_1(x)-b x\right)^2=k_1.
\end{equation}

In the classical case, $\h=0$, this is enough to solve. The
solution, which can be written implicitly in a more compact form,
is given by
\begin{equation}
d=V_1(V_1-b x)^2,
\end{equation}
\noindent where $d$ is an arbitrary constant. When $d=0$ we find
the familiar case $V=b x+a y$ which admits a first-order integral.
We may notice that the implicit form of the solution is somewhat
similar to \eqref{implic1}.

In the quantum case, we notice that the transformation
$W(x)=V_1(x)-b x$, that preserves the Painlev\'e property,
simplifies equation \eqref{intprem2} that becomes
\begin{equation}
 \h^2 \left(2 WW''-W'^2\right)-8 \left(W+b x\right)W^2=k_2.
\end{equation}
or
\begin{equation}\label{nlordre2}
W''=\frac{W'^2}{2 W}+\frac{4 W^2}{\h^2}+\frac{4 b x
W}{\h^2}+\frac{k_2}{2 \h^2 W}
\end{equation}
We can also substitute $Y(x)=± \sqrt{W(x)}$ to find
\begin{equation*}
 \h^2 Y''=2 \left(Y^2+b x\right)Y+\frac{k_2}{4 Y^3}.
\end{equation*}

If $k_2=0$, we can set $\h=1$ and $b=1/2$ by the change of variables
\begin{equation*}
Y=(2\h b)^\frac{1}{3} Z,\; x=\left(\frac{\h^2}{2b}\right)^\frac{1}{3} \xi.
\end{equation*}
The solution for $Z(\xi)$ is then a special case of the second
Painlev\'e transcendant, defined by the equation

\begin{equation*}
P_2''(x,\a)=2P_2(x,\a)^3+x P_2(x,\a)+\a.
\end{equation*}

 The solution for $V$ is
\begin{equation}
V(x,y)=b x+a y +(2\h
b)^{\frac{2}{3}}P_2^2\left(\left(\frac{2b}{\h^2}\right)^\frac{1}{3}
x,0\right),
\end{equation}

\noindent with the particular case $V=bx+ay$.

Now if $k_2\neq 0$, we use \eqref{nlordre2}, that can be
normalized by a change of variables,
$$x= -\left(\frac{\h^2}{(4 b)}\right)^\frac{1}{3}\xi,\;
W= -\frac{\sqrt{-k_2} }{(4\h b)^\frac{1}{3}} Y.$$
 We then find

\begin{equation}\label{cas34}
Y''=\frac{Y'^2}{2 Y}+4 \b Y^2- \xi Y-\frac{1}{2Y},
\end{equation}

\noindent where $\b = -\sqrt{-k_2}/(4\h b) \neq 0$ is an arbitrary
constant. Equation \eqref{cas34} corresponds to case XXXIV, p.
340, in \cite{In}. The solution for $Y(\xi)$ reads
\begin{equation}
2 \b
Y=P_2'(\xi,-2\b-\frac{1}{2})+P_2(\xi,-2\b-\frac{1}{2})^2+\frac{\xi}{2}
\end{equation}

\noindent where $P_2$ is once again the second Painlev\'e
transcendant. Since $\b$ is an arbitrary constant, we can set
$\kappa=-2\b-\frac{1}{2}$.

Back to the original variables, we get

\begin{equation}
V(x,y)=a y +\left( 2 \h^2 b^{2}\right)^\frac{1}{3} \left( P_2'(-(4
b/\h^2)^\frac{1}{3} x, \kappa)+ P_2^2(-(4 b/\h^2)^\frac{1}{3} x,
\kappa)\right)
\end{equation}

 This
potential admits $V=ay$ and $V=ay+\h^2/x^2$ as particular cases.

In all cases with $V=a y+V_1(x)$ and $A_{012}A_{030}\neq0$ the
integral of motion is
\begin{equation}
2 a p_x^3-2 b p_x^2 p_y+a\{3 V_1(x)-b x,p_x\}-2b\{V_1(x),p_y\}.
\end{equation}

Limiting values for $a$, $b$ and $\h$ give either known potentials
or trivial integrals.

\subsection{Case \ref{qcque}: $A_{120}=A_{021}=0$}

Since we set here $A_{120}=A_{021}=0$, we assume
$A_{030}A_{003}\neq 0$ (otherwise we would find potentials of the
form \eqref{Weiersol}). The conditions for the existence of a
third-order operator then read

\begin{equation}\label{transc}
\begin{split}
\h^2 V_1''(x)= 6 V_1^2(x)+A_{003} \sigma x\\
\h^2 V_2''(y)= 6 V_2^2(y)-A_{030} \sigma y.
\end{split}
\end{equation}

If $\sigma$ vanishes, we find the potential

$$V=\h^2(\mathcal{P}(x)+\mathcal{P}(y)),$$

\noindent with two integrals that each depend on only one
variable, as could have been predicted from the results of
equation \eqref{Weierstrass}. If $\sigma$ does not vanish, let us
set $b_1= A_{003} \sigma$ and $b_2=-A_{030} \sigma$.

If $\h=0$, we find that $$V=\pm \sqrt{\beta_1 x}\pm \sqrt{\beta_2
y}$$ is superintegrable, where the $\beta_i=b_i/6$ are arbitrary
constants.

This defines similar real potentials on each quadrant of the
plane. We can again patchwork the pieces to find real continuous
potentials defined everywhere, e.g. $V= c_1
\sqrt{|x|}+c_2\sqrt{|y|}$, although in that case neither the
Hamiltonian nor the integral are differentiable at the equilibrium
point.

Finally, if $\h \neq 0$, we find that $V_1$ and $V_2$ can be both
written using the first Painlev\'e transcendent.

\begin{equation}V=\h^2 \w_1^2P_1\left(\w_1 x\right)+\h^2
\w_2^2P_1\left(\w_2 x\right)\end{equation}
 \noindent where the $\w_i=(b_i/\h^4)^{1/5}$ are arbitrary constants.

The integral of motion is

\begin{equation}
X= 2 b_2 p_x^3-2 b_1 p_y^3+3 b_2 \{ V_1(x,b_1),p_x\}-3
b_1\{V_2(y,b_2),p_y\},
\end{equation}

both in quantum and classical mechanics.




\section{Conclusion}

We have found all systems in two-dimensional euclidian space that
admit separation of variables in cartesian coordinates and at
least one third-order integral, both in quantum and classical
mechanics. Many new superintegrable potentials were found, and,
interestingly, all the quantum superintegrable potential are found
as solutions of equations having the Painlev\'e property, and this
is probably not accidental. Many of the quantum potentials can in
fact be written in terms of different transcendent functions of
Painlev\'e, and many are related to group invariant solutions of
the Korteweg-de Vries equation, and to reductions of the
Boussinesq and KP equations. All classical integrable potentials
were found to be limiting cases of quantum ones. They do not obey
to equations having the Painlev\'e property, though, for many
classical integrable potentials have movable branch points of the
form $\sqrt{x-b}$. Thus, in that respect, quantum integrable
potentials behave more regularly than classical ones. A natural
question is what does the Painlev\'e property, in quantum
mechanics, tell us about classical integrable potentials.
Since this paper deals mostly with the classification of superintegrable systems,
the consideration of such questions regarding their properties and their solutions will be postponed to a future article.

Our investigation provided interesting new examples of the
differences between quantum and classical integrability. A
systematic search for systems with higher-order integrals is
therefore a useful task since we still know little about such
systems, and they are likely to share interesting properties.

Note added in proof: The future article mentioned earlier (\cite{Gr2}) indeed shows that superintegrability of separable systems tells a lot about the physical properties of the separated systems. Even though the systems considered here can all be separated in two independent one-dimensional systems, two-dimensional superintegrability provides information on them that is far from obvious from a one-dimensional perspective.

Equations \eqref{eqquant1} to \eqref{eqquant4} are the necessary
and sufficient conditions for the existence of third-order
integrals. The general solution to these equation is highly
nontrivial, and it is not likely that a direct approach will lead
to such a solution for integrals of order higher than four,
without the use of new methods. In order to develop such methods,
it would be useful to have a rigorous definition of quantum
integrability.

\section*{Acknowledgements}

The author thanks gratefully P. Winternitz for his very helpful
comments and suggestions, and also J. B\'erub\'e for useful
discussions. The author benefits from a NSERC ES-A fellowship.

\section{Annex I}

Here is the complete list of two-dimensional Hamiltonians
separable in cartesian coordinates that admit at least one
third-order integral. We also gave all their third-order
integrals. Potentials depending on only one cartesian coordinates
are not listed here, for they were classified in \cite{GW}. Many
of the potentials listed below were already known to be
superintegrable, but we listed them here for completeness and in
some cases to take into account additional third-order integrals.
Potentials in boxes are those who do not admit enough first-
or second-order integrals to make them superintegrable. Most of
them were not known before.

Some potentials (namely $V_p$ and $V_q$) are not superintegrable
according to the usual definition, for their integrals have simple
though nontrivial relations with the Hamiltonian.  For definitions
of parameters the reader is referred to the body of the article.
In order to write the integrals of motion in a compact form we
often use the notation $V(x,y)=V_{1}(x)+V_{2}(y)=V_{1}+V_{2}$.

\subsection{Quantum potentials}

\begin{enumerate}[(Q.1)]

  \item  $V=a (x^2 +y^2)\\
  X_{1}= L^3\\
  X_{2}=\{L,p_x p_y\}+a\{2 x^2 y,p_y\}-a \{2 x y^2,p_x\}\\
  X_{3}=\{L,p_y^2\}+2 a\left(\{ x y^2,p_y\}-\{y^3,p_x\}\right)\\
  X_{4}=\{L,p_x^2\}-2 a\left(\{ x^2
  y,p_x\}-\{x^3,p_y\}\right)$

  \item $ V=a (x^2 +y^2)+\frac{b}{x^2}+\frac{c}{y^2}\\
  X_{1}= \{L,p_xp_y\}+\{xy\left(-\frac{2b}{x^3}+2 ax\right),p_y\}-\{xy\left(-\frac{2c}{y^3}+2 a
  y\right),p_x\}$

  \item $ V=a (x^2 +y^2)+\frac{\h^2}{x^2}+\frac{\h^2}{y^2}\\
  X_{1}=2 L^3 - \h^2 \{\frac{3x^2}{y}+2 y+\frac{3y^3}{x^2},p_x\}+ \h^2 \{\frac{3 y^2}{x}+2x+\frac{3 x^3}{y^2},p_y\}\\
  X_{2}= \{L,p_xp_y\}+\{xy\left(-\frac{2\h^2}{x^3}+2 ax\right),p_y\}-\{xy\left(-\frac{2\h^2}{y^3}+2 a
  y\right),p_x\}$

  \item $ V=a (x^2 +y^2)+\frac{\h^2}{y^2}\\
  X_{1}= 2 L^3-\h^2 \{2 y+\frac{3x^2}{y},p_x\}+\h^2\{\frac{3 x^3}{y^2}+2x,p_y\}\\
  X_{2}=\{L,p_x p_y\}- 2\{ ax y^2-\frac{\h^2x}{y^2},p_x\}+2a\{x^2y,p_y\}\\
  X_{3}=\{L,p_y^2\}- \{2 ay^3+\h^2\frac{1}{y},p_x\}+\{\h^2 \frac{3
  x}{y^2}+2axy^2,p_y\}$


  \item $\boxed{V_{}=\h^2\left( \frac{1}{8\a^4}(x^2 +y^2)+\frac{1}{(x-\a)^2}+\frac{1}{(x+\a)^2}\right)}\\
  X_{1}=2L^3-3\a^2 \{L,p_y^2\}+\frac{\h^2}{4} \{y(-8+\frac{3y^2}{\a^2}-\frac{24 y^2 (x^2+\a^2)}{(x-\a)^2(x+\a)^2}),p_x\}\\
  +\frac{\h^2}{4}\{x(8-\frac{3 y^2(x^4-10 \a^2 x^2-24\a^4)}{\a^2(x-\a)^2(x+\a)^2}),p_y\}\\
  X_{2}= \{L,p_x^2\}+\h^2 \{y(\frac{4\a^2-x^2}{4\a^4}-\frac{6(x^2+\a^2))}{(x^2-\a^2)^2}),p_x\}\\
  \;\;\;\;+\h^2\{\frac{x(x^2-4\a^2)}{4 \a^4}-\frac{2x}{x^2-\a^2}+\frac{4 x (x^2+\a^2)}{(x-\a)^2(x+\a)^2},p_y\}$


\item $  \boxed{V_{}=\h^2\left( \frac{1}{8 \a^4}(x^2 +y^2)+\frac{1}{y^2}+\frac{1}{(x+\a)^2}+\frac{1}{(x-\a)^2}\right)}\\
  X_{1}=2 L^3-3\a^2\{L,p_y^2\}+\h^2\{\frac{3y^3}{4\a^2}+\frac{6y^3(x^2+\a^2)}{(x-\a)^2(x+\a)^2}-\frac{3(x^2-\a^2)}{y}-2y,p_x\}
+3\h^2\{x(\frac{x^2-3\a^2}{y^2}-\frac{3y^2-8\a^2}{12\a^2}-\frac{2
y^2 }{x^2-\a^2}+\frac{4 y^2(x^2+\a^2)}{(x-\a)^2(x+\a)^2}),p_y\}$


\item  $ \boxed{V_{}=\h^2\left( \frac{1}{8 \a^4}(x^2 +y^2)+\frac{1}{(y-\a)^2}+\frac{1}{(x-\a)^2}+\frac{1}{(y+\a)^2}+\frac{1}{(x+\a)^2}\right)}\\
X_{1}=2
L^3-3\a^2(\{L,p_x^2\}+\{L,p_y^2\})+\frac{\h^2}{4}\{y(124+\frac{3(x^2+y^2)}{\a^2}+\frac{24(x^2-5y^2)}{y^2-\a^2}-\frac{144x^2}{x^2-\a^2}
+\frac{24(3x^2-y^2)(x^2+\a^2)}{(x-\a)^2(x+\a)^2}+\frac{48(y^2-x^2)(y^2+\a^2)}{(y-\a)^2(y+\a)^2}),p_x\}-\frac{\h^2}{4}\{x(124+\frac{3(x^2+y^2)}{\a^2}-\frac{24(5x^2-y^2)}{x^2-\a^2}-\frac{144x^2}{y^2-\a^2}
-\frac{24(x^2-3y^2)(y^2+\a^2)}{(y-\a)^2(y+\a)^2}+\frac{48(x^2-y^2)(x^2+\a^2)}{(x-\a)^2(x+\a)^2}),p_y\}$


  \item $  V=a (4 x^2 +y^2)+\frac{b}{y^2}+c x\\
  X_{1}= 2 p_x p_y^2+\{-2 ay^2+\frac{2 b}{y^2},p_x\}+ \{ 8a
  xy+cy,p_y\}$

  \item $  \boxed{V=a (9 x^2 +y^2)}\\
  X_{1}=\{L,p_y^2\}+\frac{2}{3} a\{y^3,p_x\}-6 a\{ x
  y^2,p_y\}$

  \item $ \boxed{V=a (9 x^2 +y^2)+\frac{\h^2}{y^2}}\\
   X_{1}=\{L,p_y^2\}+\{\frac{2 a y^3}{3}-\frac{\h^2}{y},p_x\}+\{3 x\left(-2 a y^2+\frac{\h^2}{y^2}\right),p_y\}$


  \item $ \boxed{V_{}=\h^2\left( \frac{1}{8\a^4}(9 x^2 +y^2)+\frac{1}{(y+\a)^2}+\frac{1}{(y-\a)^2}\right)}\\
  X_{1}=\{L,p_y^2\}+\h^2\{y\left(\frac{y^2}{12\a^4}-\frac{8\a^2}{(y^2-\a^2)^2}-\frac{2}{y^2-\a^2}\right),p_x\}+\frac{3\h^2}{4}\{x(\frac{8(y^2+\a^2)}{(y^2-\a^2)^2}-\frac{y^2}{\a^4}),p_y\}$

  \item $ \boxed{V=\frac{\h^2}{x^2}+\frac{a}{y^2}}\\
  X_{1}=\{L^2,p_x\}-2\h^2\{ \frac{y^2}{x^2},p_y\}+\{3\h^2
  \frac{y^2}{x^2}+2a \frac{x^2}{y^2}+\frac{\h^2}{2},p_x\}\\
  X_{2}= \{L,p_xp_y\}-2\h^2\{\frac{y}{x^2},p_y\}+2a\{\frac{x}{y^2},p_x\}\\
  X_{3}=2 p_x^3+\{\frac{3 \h^2}{x^2},p_x\}$

  \item $ \boxed{V=\frac{\h^2}{x^2}+\frac{\h^2}{y^2}}\\
  X_{1}=2 L^3-\h^2\{2y+\frac{3x^2}{y}+\frac{3y^3}{x^2},p_x\}+\h^2\{2x+\frac{3y^2}{x}+\frac{3x^3}{y^2},p_y\}\\
  X_{2}=\{L^2,p_x\}-2\h^2\{ \frac{y}{x},p_y\}+\{3\h^2
  \frac{y^2}{x^2}+2\h^2 \frac{x^2}{y^2}+\frac{\h^2}{2},p_x\}\\
  X_{3}=\{L^2,p_y\}-2\h^2\{ \frac{x}{y},p_x\}+\{3\h^2
  \frac{x^2}{y^2}+2\h^2 \frac{y^2}{x^2}+\frac{\h^2}{2},p_x\}\\
  X_{4}= \{L,p_xp_y\}-2\h^2\{\frac{y}{x^2},p_y\}+2\h^2\{\frac{x}{y^2},p_x\}\\
  X_{5}=2 p_x^3+\{\frac{3 \h^2}{x^2},p_x\}\\
  X_{6}=2 p_y^3+\{\frac{3\h^2}{y^2},p_y\}$

\item $  V= a x +\frac{\h^2}{y^2}\\
  X_{1}=\{L,p_y^2\}-\{\frac{\h^2}{y},p_x\}+\{3\frac{\h^2 x}{y^2}-a\frac{y^2}{2},p_y\}\\
  X_{2}= 2 p_y^3+\{3 \frac{\h^2}{y^2},p_y\}\\
  X_{3}=2p_xp_y^2 +2 \h^2\{\frac{1}{y^2},p_x\}+a\{y,p_y\}$

%

\item $ \boxed{V=\h^2 \mathcal{P} (y)+V(x)}\\
  X_{1}=2 p_y^3+\{3 \h^2 \mathcal{P} (y),p_y\}$

\item $ \boxed{V_{}=\h^2(\mathcal{P}(x)+\mathcal{P}(y))}\\
X_{1}=2 p_x^3+\{3 \h^2 \mathcal{P} (x),p_x\}\\
X_{2}=2 p_y^3+\{3 \h^2 \mathcal{P} (y),p_y\}$

\item
$\boxed{V_{}=\h^2 \w_1^2P_1\left(\w_1 x\right)+\h^2
\w_2^2P_1\left(\w_2 x\right)}\\
X_{1}= 2 \w^5_2 p_x^3-2 \w^5_1 p_y^3+3 \w^5_2 \{
V_{1}(x,\w_1),p_x\}-3 \w^5_1\{V_{2}(y,\w_2),p_y\}$

\item $ \boxed{V=a (x^2+y^2)+\frac{\h}{2}b_1 P_4'(x,\frac{-8a}{\h^2})\\+4a
P_4^2(x,\frac{-8a}{\h^2})+4 a x P_4(x,\frac{-8
a}{\h^2})+\frac{1}{6}(-\h^2 K_1+\h b_1)}\\\\
X_{1}=\{L,p_x^2\}+\{ax^2y-3y
V_{1},p_x\}-\frac{1}{2a}\{\frac{\h^2}{4}V_{1xxx}+(ax^2-3V_{1})
V_{1x} ,p_y\}$

\item
$ \boxed{V=a y+\h^2 \w^2 P_1(\w x)}\\
X_{1}=2 p_x^3+3 \h^2 \w^2 \{P_1(\w x),p_x\}+\{\frac{\w^5\h^4}{4
a}, p_y\}$

\item $ \boxed{V=b x+a y +(2\h
b)^{\frac{2}{3}}P_2^2\left(\left(\frac{2b}{\h^2}\right)^\frac{1}{3}
x,0\right)}\\
X_{1}=2 a p_x^3-2 b p_x^2 p_y+a\{3V_{1}(x)-b
x,p_x\}-2b\{V_{1}(x),p_y\}$

\item $ \boxed{V=a y +\left( 2 \h^2 b^{2}\right)^\frac{1}{3} \left(
P_2'(-(4 b/\h^2)^\frac{1}{3} x, \kappa)+ P_2^2(-(4
b/\h^2)^\frac{1}{3} x, \kappa)\right)}\\
X_{1}=2 a p_x^3-2 b p_x^2 p_y+a\{3V_{1}(x)-b
x,p_x\}-2b\{V_{1}(x),p_y\}$
\end{enumerate}

\subsection{Classical potentials}

\begin{enumerate}[(C.1)]

 \item $ V=a (x^2 +y^2)\\
  X_{1}= L^3\\
  X_{2}=\{L,p_x p_y\}+a\{2 x^2 y,p_y\}-a \{2 x y^2,p_x\}\\
  X_{3}=\{L,p_y^2\}+2 a\left(\{ x y^2,p_y\}-\{y^3,p_x\}\right)\\
  X_{4}=\{L,p_x^2\}-2 a\left(\{ x^2
  y,p_x\}-\{x^3,p_y\}\right)$

  \item $ V=a (x^2 +y^2)+\frac{b}{x^2}+\frac{c}{y^2}\\
  X_{1}= \{L,p_xp_y\}+\{xy\left(-\frac{2b}{x^3}+2 ax\right),p_y\}-\{xy\left(-\frac{2c}{y^3}+2 a
  y\right),p_x\}$

 \item $  V=a (4 x^2 +y^2)+\frac{b}{y^2}+c x\\
  X_{1}= 2 p_x p_y^2+\{-2 ay^2+\frac{2 b}{y^2},p_x\}+ \{ 8a
  xy+cy,p_y\}$

  \item $  \boxed{V=a (9 x^2 +y^2)}\\
  X_{1}=\{L,p_y^2\}+\frac{2}{3} a\{y^3,p_x\}-6 a\{ x
  y^2,p_y\}$

\item $ \boxed{V_{}=\pm \sqrt{\beta_1 x}\pm \sqrt{\beta_2 y}}\\
X_1= 2 \b_2 p_x^3-2 \b_1 p_y^3+3 \b_2 \{ V_{1}(x,6 \b_1),p_x\}-3
\b_1\{V_{2}(y,6\ b_2),p_y\}$

\item $ \boxed{V_{}= ay^2 +V_{1}}$ \:\: where $V_{1}$ satisfies equation \eqref{implic1}\\
$ X_{1}=\{L,p_x^2\}+\{ax^2y-3y
V_{1},p_x\}-\frac{1}{2a}\{(ax^2-3V_{1}) V_{1x} ,p_y\}$

\item $ \boxed{V_{}=ay+b\sqrt{x}}\\
X_{1}=2 p_x^3+3 b \{\sqrt{x},p_x\}-\{\frac{3 b^2}{2 a}, p_y\}$

\item $ \boxed{V_{}=a y+ V_{1}(x)}$,\;\;\:where
$(V_{1}-bx)^2 V_{1}=d$\\ $X_{1}=2 a p_x^3-2 b p_x^2
p_y+a\{3V_{1}(x)-b x,p_x\}-2b\{V_{1}(x),p_y\}$

\end{enumerate}


\end{document}